\documentclass[twocolumn,showpacs,preprintnumbers,prc]{revtex4-1}
\usepackage{graphicx,color}
\usepackage{float}
\usepackage{bm}
\begin{document}

\title{Statistical Hauser-Feshbach theory with width fluctuation
       correction including direct reaction channels
       for neutron induced reaction at low energies}

\author{T. Kawano}
\email{kawano@lanl.gov}
\affiliation{Theoretical Division,
  Los Alamos National Laboratory, Los Alamos, NM 87545, USA}

\author{R. Capote}
\affiliation{NAPC--Nuclear Data Section,
  International Atomic Energy Agency, Vienna A-1400, Austria}

\author{S. Hilaire}
\affiliation{CEA/DIF,
  Service de Physique Nucl\'{e}aire,
  F-91680 Bruy\`{e}res-le-Ch\^{a}tel, France}

\date{\today}

\begin{abstract}
  A model to calculate particle-induced reaction cross sections with
  statistical Hauser-Feshbach theory including direct reactions is
  given. The energy average of scattering matrix from the
  coupled-channels optical model is diagonalized by the transformation
  proposed by Engelbrecht and Weidenm\"{u}ller. The ensemble average
  of $S$-matrix elements in the diagonalized channel space is
  approximated by a model of Moldauer [Phys.Rev.C {\bf 12}, 744
    (1975)] using newly parametrized channel degree-of-freedom $\nu_a$
  to better describe the Gaussian Orthogonal Ensemble (GOE) reference
  calculations.  Moldauer approximation is confirmed by a Monte Carlo
  study using randomly generated $S$-matrix, as well as the GOE
  three-fold integration formula. The method proposed is applied to
  the $^{238}$U(n,n') cross section calculation in the fast energy
  range, showing an enhancement in the inelastic scattering cross
  sections.
\end{abstract}
\maketitle

\section{Introduction}
\label{sec:introduction}

Neutron scattering in the keV to MeV energy range is one of the most
important processes in many fields, for which better understanding of
nuclear reaction mechanisms is always crucial. In particular, accurate
neutron reaction cross sections are needed for applications such as
radiation transport simulations for nuclear technology, particle
detector response, nuclear reaction rate calculation for nuclear
astrophysics, and so forth. When we calculate the nuclear reaction
cross section for a system where the dynamical or static nuclear
deformation is involved, the simple regime of the spherical optical
model plus the Hauser-Feshbach theory \cite{Hauser52} has to be
extended to the coupled-channels scheme
(e.g. Ref.~\cite{Tamura}). Rotational bands built on intrinsic or
vibrational levels dominate the low-lying excitation spectra for
statically deformed nuclei, and it is well known that these excited
rotational states are strongly populated by the collective motion of
target nucleus.

Typically, the direct reaction channels in the statistical model have
been considered in a perturbed way, in which a flux going into the
direct channels is subtracted from the total compound nucleus
formation cross section \cite{Kawano09}, i.e., the direct and compound
cross sections are assumed to be independent.  Such approximation has
a great advantage to reduce computational burden, and therefore, many
Hauser-Feshbach codes, such as Empire \cite{Empire},
TALYS \cite{TALYS}, CCONE \cite{Iwamoto07}, CoH$_3$ \cite{CoH3, Kawano10},
etc., employ this approximation to calculate nuclear reaction cross
sections. However,it was shown that the existence of direct reaction
channels changes the compound reaction cross
sections \cite{Kawano15b}. Therefore it is important to assess the
independence of the direct and compound reaction mechanisms
quantitatively, which exists implicitly in the approximation
aforementioned.

Statistical models for the compound nuclear reaction connect energy
average $S$-matrix elements (or transmission coefficients) to energy
average cross sections. While the statistical Hauser-Feshbach
theory provides such a link, it has to be modified by the width
fluctuation correction that accounts for statistical properties in the
resonances. The width fluctuation correction enhances the cross
section in the elastic channel, and reduces all other channels to
fulfill the unitarity condition.  When strongly coupled channels
exist, the energy average $S$-matrix, $\langle S\rangle$, is no-longer
diagonal.  The imposed unitarity condition yields additional
correlations between the elastic and other channels, hence the cross
sections will be further modified \cite{Kawai73}.

Kawai, Kerman, and McVoy (KKM) \cite{Kawai73} obtained a formula for
the compound nuclear reaction including the direct channels at the
strong absorption limit.  The actual calculations of KKM are,
unfortunately, very limited \cite{Arbanas08, Kawano08}.  In parallel
to KKM, inclusion of the direct reaction in the statistical theory was
proposed by Engelbrecht and Weidenm\"{u}ller \cite{Engelbrecht73}, in
which $\langle S\rangle$ is diagonalized by a unitary transformation.
The statistical model calculation is performed in the diagonalized
space, just like the no-direct reaction cases. Hofmann et
al. \cite{Hofmann75} and Moldauer \cite{Moldauer75b} performed the
Engelbrecht-Weidenm\"{u}ller (EW) transformation to examine the
effects of the direct channels on the compound nuclear reaction. A
more general and rigorous theory was proposed by Nishioka,
Weidenm\"{u}ller, and Yoshida (NWY) \cite{Nishioka89} based on the
so-called Gaussian Orthogonal Ensemble (GOE) \cite{Verbaarschot85}
together with the EW transformation. However, the NWY equation
obtained is almost impossible to calculate. The most recent study on
this subject is by Capote et al. \cite{Capote14}, who studied the
impact of the EW transformation on a realistic calculation of
inelastic scattering on $^{238}$U using the coupled-channels optical
model code ECIS \cite{ECIS}. An enhancement of the inelastic
scattering cross section was found \cite{Capote14}, yet the compound
reaction model implemented in ECIS is limited and further
investigation was needed.

In the case of a spherical nucleus, we obtained a simple relationship
between the channel degree-of-freedom $\nu_a$ and the optical model
transmission coefficients $T_a$ by applying the Monte Carlo technique
to GOE \cite{Kawano14}, which yields an almost equivalent compound
nucleus cross sections to the GOE three-fold integration
formula \cite{Verbaarschot85}.  Such an empirical approach facilitates
computations of the Hauser-Feshbach theory in the fast energy range,
where the number of open channels tends to be too large to handle.
Starting with the approach by Moldauer \cite{Moldauer75b}, and adding
the idea of GOE three-fold integration, we extend Moldauer's approach
to the actual cross section calculation for deformed nuclei.  Since we
will show in this paper that our model produces almost identical
results to the NWY theory, the calculated nuclear reaction cross
sections should be within reasonable uncertainties for many realistic
cases.  This could be particularly important to calculate nuclear
reaction cross sections for actinides or in the rare earth region,
where the static nuclear deformation is large.

\section{Theory}
\label{sec:theory}

\def\ave#1{\left\langle{#1}\right\rangle}

\subsection{Hauser-Feshbach theory with width fluctuation correction}

In the case of nuclear reaction without direct channels, the
Hauser-Feshbach theory with the width fluctuation correction reads
\begin{equation}
  \sigma_{ab}
  = \frac{\pi}{k^2_a} \frac{T_a T_b}{\sum_c T_c} W_{ab}
  = \sigma_{ab}^{\rm HF} W_{ab} ,
  \label{eq:HF}
\end{equation}
where $\sigma_{ab}$ is the energy average cross section from channel
$a$ to $b$, $\sigma_{ab}^{\rm HF}$ is the Hauser-Feshbach cross
section, $k_a$ is the wave-number of projectile, $W_{ab}$ is the width
fluctuation correction factor, and $T_c$ is the transmission
coefficient in channel $c$ calculated with the optical model
$S$-matrix element $T_c = 1 - |\ave{S_{cc}}|^2$.  Hereafter we omit
the kinematic factor of $\pi/k_a^2$, unless otherwise specified.

The width fluctuation correction factor is given by the Gaussian
Orthogonal Ensemble (GOE) model of Verbaarschot, Weidenm\"{u}ller, and
Zirnbauer \cite{Verbaarschot85}. This model gives an ensemble average
of the fluctuation part, $\overline{S_{ab} S_{cd}^*}$, and the width
fluctuation correction factor can be calculated as a ratio to $\sigma_{ab}^{\rm HF}$.
The so-called GOE triple-integral formula is \cite{Verbaarschot85}
\begin{widetext}
\begin{equation}
 \overline{S_{ab} S_{cd}^*} =
    \frac{1}{8}
    \int_0^\infty d\lambda_1 \int_0^\infty d\lambda_2  \int_0^1 d\lambda  \quad
    \mu(\lambda,\lambda_1,\lambda_2) \nonumber\\
    {\displaystyle \prod_c}
    \frac{1 - T_c\lambda }{\sqrt{(1+T_c\lambda_1)(1+T_c\lambda_2)}}
    J(\lambda,\lambda_1,\lambda_2)  ,
   \label{eq:GOE3int}
\end{equation}
where
\begin{eqnarray}
\mu(\lambda,\lambda_1,\lambda_2)
  &=&  \frac{\lambda(1-\lambda) | \lambda_1 - \lambda_2 |}
       {\sqrt{\lambda_1(1+\lambda_1)} \sqrt{\lambda_2(1+\lambda_2)}
        (\lambda+\lambda_1)^2 (\lambda+\lambda_2)^2 }  ,  \\
  J(\lambda,\lambda_1,\lambda_2)
  &=& \delta_{ab} \delta_{cd}
      \overline{S}_{aa} \overline{S}_{cc}^* T_a T_c
      \left(
        \frac{ \lambda_1}{1+T_a\lambda_1}
      + \frac{ \lambda_2}{1+T_a\lambda_2}
      + \frac{2\lambda  }{1-T_a\lambda  }
      \right)
      \left(
        \frac{ \lambda_1}{1+T_c\lambda_1}
      + \frac{ \lambda_2}{1+T_c\lambda_2}
      + \frac{2\lambda  }{1-T_c\lambda  }
      \right) \nonumber \\
  &+& (\delta_{ac}\delta_{bd} + \delta_{ad}\delta_{bc}) T_a T_b
      \left\{
         \frac{ \lambda_1(1+\lambda_1)}{ (1+T_a\lambda_1) (1+T_b\lambda_1) }
       + \frac{ \lambda_2(1+\lambda_2)}{ (1+T_a\lambda_2) (1+T_b\lambda_2) }
       + \frac{2\lambda  (1-\lambda  )}{ (1-T_a\lambda  ) (1-T_b\lambda  ) }
     \right\}  .
     \label{eq:GOEJ}
\end{eqnarray}
\end{widetext}

The compound cross section is readily calculated as $\overline{S_{ab}
S_{ab}^*} = \overline{|S_{ab}|^2} = \sigma_{ab}$ when $\ave{S}$ is
provided, beside the time-consuming three-fold
integration \cite{Verbaarschot86}.  The GOE model is believed to be a
correct answer to the calculation of the compound cross
section. However, it is not so practical to apply
Eq.~(\ref{eq:GOE3int}) to realistic cases.  For example, a compound
nucleus after a particle or photon emission is often left in the
continuum state, where the decay channel is not well defined. Even if
we approximate the transition to one of the continuum bins by a
pseudo-single level, the calculation time will be enormous when there
are many open channels. Alternatively, there are several models to
evaluate $W_{ab}$. We adopt Moldauer's model
\cite{Moldauer75a, Moldauer75b, Moldauer76, Moldauer78}, since Hilaire,
Lagrange, and Koning \cite{Hilaire03} reported that this model is
practically accurate enough. The width fluctuation correction factor
can be evaluated numerically as
\begin{eqnarray}
  W_{ab} &=& \left( 1+\frac{2\delta_{ab}}{\nu_a} \right)
   \int_0^\infty\!\!\!
   \frac{dt}{F_a(t)F_b(t) \prod_k F_k(t)^{\nu_k/2}} ,
     \label{eq:MoldauerW}  \\
  F_k(t) &=& 1 + \frac{2}{\nu_k} \frac{T_k}{\sum_c T_c} t ,
\end{eqnarray}
where $\nu_a$ is the channel degree-of-freedom, which is related to
the channel transmission coefficient $T_a$. There are, again, several
models to express $\nu_a$ by $T_a$, which were derived by a Monte
Carlo study, such as that of Moldauer \cite{Moldauer80}, Ernebjerg and
Herman \cite{Ernebjerg04}, or of LANL \cite{Kawano14}. We here
employ the most recent model from LANL \cite{Kawano14}, because it
produces almost identical $W_{ab}$ compared to the GOE triple-integral
calculation \cite{Kawano15b}.

\subsection{Generalized transmission coefficient}

When direct reaction channels exist, in other words, the optical model
$S$-matrix is not diagonal, the Hauser-Feshbach cross section in
Eq.~(\ref{eq:HF}) should be further modified. In this case the energy
average $S$-matrix is given by the coupled-channels calculation.  When
combining the coupled-channels method with the Hauser-Feshbach theory,
the existing cross section calculation codes, such as
Empire \cite{Empire}, TALYS \cite{TALYS}, CCONE \cite{Iwamoto07}, and
CoH$_3$ \cite{CoH3}, adopt a ``direct cross section eliminated''
transmission coefficient.  This is defined as the probability of
formation of compound nucleus on the $n$-th state by a nucleon having
the orbital angular momentum and spin of $l,j$:
\begin{equation}
  T_{lj}^{(n)} =
    \sum_{J\Pi} \sum_{c}
    g_{Jc}
    \left(
      1- \sum_{c'} |\ave{S_{cc'}^{J\Pi}}|^2
    \right)
    \delta_{n_c,n} \delta_{l_c,l} \delta_{j_c,j} ,
 \label{eq:modtrans}
\end{equation}
where the suffix $c$ indicates the quantum number in the channel,
$J\Pi$ is the total spin and parity, and $g_{Jc}$ is the spin factor
\begin{equation}
  g_{Jc} = {{2J+1}\over{(2j_c+1})(2I_c+1)} .
 \label{eq:gc}
\end{equation}
$I_c$ is the spin of the nucleus state.
Equation~(\ref{eq:modtrans}) gives a partial-wave contribution to the
total compound formation cross section when the target is in its $n$-th state
\begin{equation}
 \sigma^{{\rm CN}(n)}
  = {{\pi}\over{k_n^2}} \sum_{lj}  {{2j+1}\over{2s+1}} T_{lj}^{(n)}  ,
 \label{eq:sigCN}
\end{equation}
where $s$ is the intrinsic spin of incoming particle.
Because we eliminate the off-diagonal elements in $\ave{S}$ by
Eq.~(\ref{eq:modtrans}), the meaning of the transmission coefficient
is different from the no-direct reaction case. We call this a
{\it generalized transmission coefficient.}

The statistical model calculation is performed in the direct cross
section eliminated space, assuming the channels are diagonal.
Such assumption implies that the direct and compound
cross sections are independent, and the unitarity condition is fulfilled
only for the total reaction cross section. Therefore the scattering
cross sections are given by an incoherent sum of the direct and
compound components. For example, the inelastic scattering
cross section is written as
\begin{equation}
  \sigma_{ab} = \sigma_{ab}^{{\rm DI}} +
  \frac{T'_a T'_b}{\sum_c T'_c} W_{ab} ,
  \label{eq:HFmodT}
\end{equation}
where the direct cross section $\sigma_{ab}^{{\rm DI}}$ is usually
given by the coupled-channels calculation, and we denote the
generalized transmission coefficients by $T'$. Often another
approximation is made in addition to Eq.~(\ref{eq:modtrans}),
which consists in replacing the decay channel transmission
coefficients $T_{lj}^{(n)}$ by
the ground state $T_{lj}^{(0)}$ calculated at a shifted energy,
$T_{lj}^{(n)}(E) = T_{lj}^{(0)}(E-E_x^{(n)})$, where $E_x^{(0)}$ is the
excitation energy of $n$-th level. This is not the case in our
study. Making use of the time-reversal property of $S$-matrix, the
transmission coefficients for each $n$-th state can be calculated
automatically by Eq.~(\ref{eq:modtrans}). Note that the impact of
this approximation is small when the optical potential depends
weakly on the incident energy.

\subsection{Engelbrecht-Weidenm\"uller transformation}

A rigorous treatment of off-diagonal elements in $\ave{S}$ is to
perform the Engelbrecht-Weidenm\"uller (EW)
transformation \cite{Engelbrecht73}. The particle penetration is
expressed in terms of Satchler's transmission matrix \cite{Satchler63}
\begin{equation}
  P_{ab} = \delta_{ab} - \sum_c \ave{S_{ac}}\ave{S_{bc}^*} ,
  \label{eq:Pmatrix}
\end{equation}
where the $S$-matrix elements $\ave{S_{ab}}$ are usually given by the
coupled-channels calculation. Since $P$ is Hermitian, this can be
diagonalized by a unitary transformation \cite{Engelbrecht73}
\begin{equation}
  (UPU^\dag)_{\alpha\beta} = \delta_{\alpha\beta} p_\alpha,
  \qquad 0 \le p_\alpha \le 1 ,
\end{equation}
and the same matrix $U$ diagonalizes the scattering matrix, i.e., 
\begin{equation}
  \ave{\tilde{S}} = U \ave{S} U^T .
\end{equation}
We use Greek subscripts for channel indices in the diagonalized space, 
and Latin subscripts for the normal space.

Since $\ave{\tilde{S}}$ is diagonal, a new transmission coefficient
in the diagonal channel space is defined as
\begin{equation}
  T_\alpha = 1 - \left|\ave{\tilde{S}_{\alpha\alpha}}\right|^2  = p_\alpha ,
  \label{eq:eigenvalue}
\end{equation}
and the statistical model calculation is performed in the diagonal
channel space to evaluate the fluctuating part
$\ave{\tilde{S}_{\alpha\beta}\tilde{S}_{\gamma\delta}^*}$. Finally
a back-transformation from the channel space to the cross-section
space reads
\begin{equation}
  \sigma_{ab}
  = \sum_{\alpha\beta\gamma\delta} U_{\alpha a}^* U_{\beta b}^* U_{\gamma a} U_{\delta b}
    \ave{ \tilde{S}_{\alpha\beta} \tilde{S}_{\gamma\delta}^* } .
  \label{eq:backtrans1}
\end{equation}

Nishioka, Weidenm\"{u}ller, and Yoshida (NWY) \cite{Nishioka89}
obtained an equivalent formula for the fluctuation cross section,
which expressed in terms of the non-diagonal $\ave{S}$. Although NWY
does not require the $P$-matrix diagonalization, a hefty computational
burden is still involved. Instead of calculating NWY, we follow the
procedure given above: the EW transformation is applied to
non-diagonal $\ave{S}$, and the GOE triple-integral of
Eq.~(\ref{eq:GOE3int}) is applied to the diagonalized channel
space. This is the most accurate procedure to calculate the cross
sections when $\ave{S}$ is not diagonal, and we consider this is the
reference GOE cross section, as this is equivalent to NWY.  Based on
this, we further develop a technique, which is feasible in realistic
cross section calculation cases, yet yields practically the same
results to the reference GOE. We follow Moldauer's
prescription \cite{Moldauer75b}, in which the
Engelbrecht-Weidenm\"uller (EW) transformation \cite{Engelbrecht73} is
invoked, although an approximation --- the decay amplitudes are
normally distributed and their real and imaginary parts are
uncorrelated --- was made to cross sections in the diagonalized space.

The back-transformation can be re-written as \cite{Hofmann75},
\begin{eqnarray}
  \sigma_{ab}
  &=& \sum_\alpha |U_{\alpha a}|^2 |U_{\alpha b}|^2 \sigma_{\alpha\alpha} \nonumber\\
  &+& \sum_{\alpha\neq\beta} U_{\alpha a}^* U_{\beta b}^*
       \left(
         U_{\alpha a} U_{\beta b} + U_{\beta a} U_{\alpha b}
       \right) \sigma_{\alpha\beta} \nonumber\\
  &+& \sum_{\alpha\neq\beta} U_{\alpha a}^* U_{\alpha b}^* U_{\beta a} U_{\beta b}
         \ave{\tilde{S}_{\alpha\alpha} \tilde{S}_{\beta\beta}^*} ,
  \label{eq:backtrans2}
\end{eqnarray}
where $\sigma_{\alpha\beta}$ is a width fluctuation corrected cross
section in the diagonalized channel space,
\begin{equation}
  \sigma_{\alpha\beta} = \frac{p_\alpha p_\beta}{\sum_\gamma p_\gamma} W_{\alpha\beta} .
  \label{eq:channelsig}
\end{equation}

Replacing the energy average (angle-bracket) by the ensemble average
(overline), the GOE triple-integral formula gives a new term of
$\ave{\tilde{S}_{\alpha\alpha} \tilde{S}_{\beta\beta}^*}$
in Eq.~(\ref{eq:backtrans2}) by setting $a=b=\alpha$ and
$c=d=\beta$. Moldauer \cite{Moldauer75b} estimated this in terms
of the channel degree-of-freedom $\nu_a$ and the width fluctuation
corrected cross section $\sigma_{\alpha\beta}$ as
\begin{equation}
  \overline{\tilde{S}_{\alpha\alpha} \tilde{S}_{\beta\beta}^*}
  \simeq \left( \frac{2}{\nu_\alpha} - 1 \right)^{1/2}
    \left( \frac{2}{\nu_\beta} - 1 \right)^{1/2}
    \sigma_{\alpha\beta} .
  \label{eq:Sab}
\end{equation}
This estimation was partially confirmed by a GOE Monte Carlo study
\cite{Kawano15a}, when $\overline{\tilde{S}_{\alpha\alpha} \tilde{S}_{\beta\beta}^*}$
is real. We generalize this expression by expanding to the case of
complex $\overline{\tilde{S}_{\alpha\alpha} \tilde{S}_{\beta\beta}^*}$.
The Jacobian of Eq.~(\ref{eq:GOEJ}) for $a=b=\alpha$ and $c=d=\beta$,
\begin{equation}
  J \propto \overline{S}_{\alpha\alpha} \overline{S}_{\beta\beta}^* T_\alpha T_\beta ,
\end{equation}
is real when ${\rm Im}(\overline{S}_{\alpha\alpha}\overline{S}_{\beta\beta})=0$.
This requires an extra phase factor as
\begin{equation}
  \overline{\tilde{S}_{\alpha\alpha} \tilde{S}_{\beta\beta}^*}
  \simeq e^{i(\phi_\alpha - \phi_\beta)}
    \left( \frac{2}{\nu_\alpha} - 1 \right)^{1/2}
    \left( \frac{2}{\nu_\beta} - 1 \right)^{1/2}
    \sigma_{\alpha\beta}  ,\\
  \label{eq:Sab2}
\end{equation}
where $\phi_\alpha =\tan^{-1} \tilde{S}_{\alpha\alpha}$.

\subsection{Decay to uncoupled states}

Actual cross section calculations involve many uncoupled or very
weakly coupled states, such as the neutron emission to the continuum,
the photon emission in the neutron radiative capture process, and
nuclear fission. In the generalized transmission calculation scheme,
inclusion of these channels is straightforward; the denominator of
Eq.~(\ref{eq:HFmodT}), ${\sum_c T'_c}$, includes the transmission
coefficients for all uncoupled channels. The particle emission
transmission coefficients may be given by the optical model,
the photon channel is calculated with the Giant Dipole Resonance (GDR)
model, etc.

In the case of EW transformation, the penetration matrix may have two
blocks
\begin{equation}
  P = \left(
  \begin{array}{cc}
    P_1 &  \\
        & P_2       \\
  \end{array} \right),
\end{equation}
where $P_1$ is the coupled channels $P$ matrix, and $P_2$ is the
diagonal part that accounts for decaying into the uncoupled states.
The unitary transformation is performed to $P_1$ only,
and the summation in the denominator of
$\sigma_{\alpha\beta}$ in Eq.~(\ref{eq:channelsig}) runs over
both the eigenvalues of $P_1$ and the diagonal elements of $P_2$.
Finally the uncoupled cross section is calculated by
\begin{equation}
  \sigma_{ab} = \sum_{\alpha} |U_{\alpha a}|^2 \sigma_{\alpha\beta} \delta_{\beta b}.
  \label{eq:uncoupled}
\end{equation}

\subsection{Monte Carlo technique for sampling $S$-matrix}

The aim of this paper is twofold; (a) understanding the limitation of
generalized transmission coefficient in Eq.~(\ref{eq:modtrans}), in
which no diagonalization procedure is required, and (b) when the
diagonalization is essential, how accurate the approximation of
Eq.~(\ref{eq:Sab2}) will be. To this end, we have to explore a large
parameter space spanning over various $S$-matrix elements and the
number of channels $\Lambda$. A natural approach is to employ the
Monte Carlo technique, which facilitates model comparisons in a large
multi-parametric space. We draw a diagonal element of $S$-matrix from
a uniform distribution inside the unit circle on the complex
plane. The diagonal elements are generated by
\begin{eqnarray}
 \ave{S_{aa}} = e^{i\phi}\sqrt{1-T_a} , \qquad 1 \le a \le \Lambda ,
 \label{eq:randomS}
\end{eqnarray}
where $0 \leq \phi < 2\pi$ and $ 0 < \sqrt{1-T_a} < 1$ are the sampled phase
and transmission coefficient from the uniform distribution. For the
off-diagonal elements, we impose another condition of
$|\ave{S_{ab}}|^2 < 0.5 |\ave{S_{aa}}||\ave{S_{bb}}|$.  The sampled
$S$-matrix is converted into $P$, and the matrix is diagonalized to
obtain its eigenvalues. If negative eigenvalues emerge, we discard
this $S$, and re-sample. The constructed matrix has a dimension
of $\Lambda \times \Lambda$.

With the generated $S$-matrix, dimensionless cross sections --- total
cross section of $\sigma^{\rm T}$, shape elastic scattering
$\sigma^{\rm SE}$, direct inelastic scattering $\sigma^{\rm DI}_{ab}$,
compound formation $\sigma^{\rm CN}$ --- are calculated in a common
way,
\begin{eqnarray}
\sigma^{\rm T} &=& 2(1 - \Re\ave{S_{aa}})  ,\\
\sigma^{\rm SE} &=& |1 - \ave{S_{aa}}|^2  ,\\
\sigma^{\rm DI}_{ab} &=& |\ave{S_{ab}}|^2  ,\\
\sigma^{\rm CN} &=& 1 - |\ave{S_{aa}}|^2 = T_a ,
\end{eqnarray}
and the reaction cross section reads $\sigma^{\rm R} = \sigma^{\rm CN} +
\sum_b\sigma^{\rm DI}_{ab}$. Here we implicitly assumed that $a$ is the particle incoming
channel.  Since $|\ave{S}|^2 \leq 1$, clearly $0 \leq \sigma_T \leq
4$.  We generate several hundred of $S$-matrices for each
$\Lambda=2 \sim 7$ case.

\section{Simulation using random $S$-matrix}
\label{sec:result1}

\subsection{Simulation for Engelbrecht-Weidenm\"{u}ller transformation}

Here we compare two methods to calculate the compound cross sections.
The first method is to employ the generalized transmission coefficients
in Eq.~(\ref{eq:modtrans}). Using the randomly generated $S$-matrix
this is written simply as
\begin{equation}
   T'_a = 1 - \sum_c |\ave{S_{ac}}|^2 .
  \label{eq:Tmod}
\end{equation}
The compound reaction cross sections are defined in the direct cross
section eliminated space,
\begin{equation}
  \sigma_{ab}' = \frac{T_a' T_b'}{\sum_c T_c'} W_{ab}',
\end{equation}
where we use Eq.~(\ref{eq:GOE3int}) to calculate $W_{ab}'$.  The
second method is to perform the EW transformation. The cross section
is given by Eq.~(\ref{eq:backtrans1}), with
$\overline{\tilde{S}_{\alpha\beta} \tilde{S}_{\gamma\delta}^*}$ by
Eq.~(\ref{eq:GOE3int}). This procedure yields the correct results,
and is thus our reference GOE cross section.

\begin{figure}
  \begin{center}
    \resizebox{0.9\columnwidth}{!}{\includegraphics{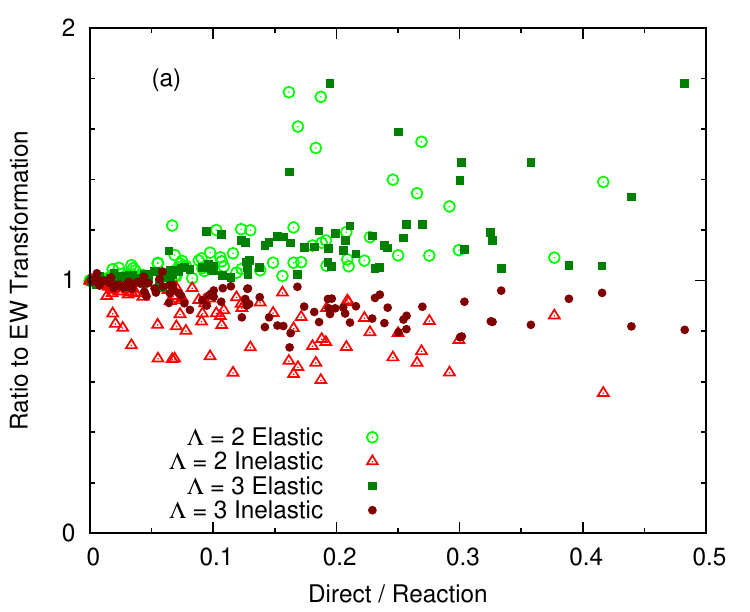}}\\
    \resizebox{0.9\columnwidth}{!}{\includegraphics{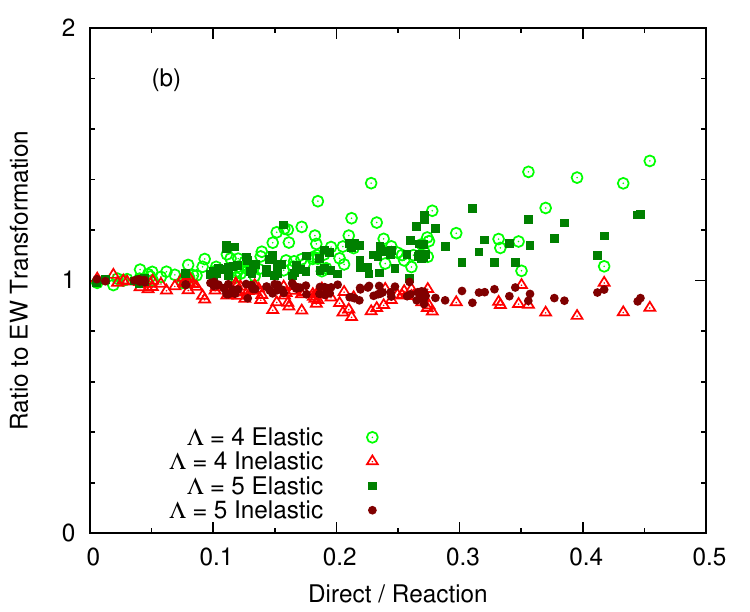}}\\
    \resizebox{0.9\columnwidth}{!}{\includegraphics{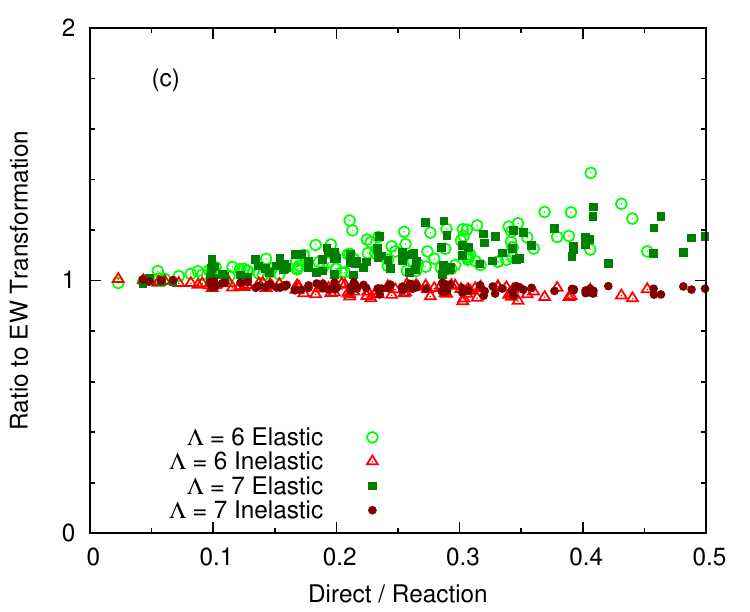}}
  \end{center}
  \caption{Ratio of calculated cross sections using randomly generated
    $S$-matrix, as a function of the direct reaction strength. The 
    ratio is that of generalized transmission coefficient calculations to the
    EW transformation case. The top panel (a) is for a number of channels of
    $\Lambda=2$ and 3, the middle panel (b) is for $\Lambda=4$ and 5,
    and the bottom panel (c) is for $\Lambda=6$ and 7.}
  \label{fig:modtrans}
\end{figure}

The calculated cross sections with the generalized transmission
coefficients are shown in Fig.~\ref{fig:modtrans} by the ratio to the
reference GOE cross sections, as a function of the strength of direct
channels $\sum_b\sigma^{\rm DI}_{ab} / \sigma^{\rm R}$ for $\Lambda=2
\sim 7$. In the case of $\Lambda > 2$, the inelastic scattering are
summed
\begin{equation}
  \sigma^{\rm INL} = \sum_{b (a \ne b)} \sigma_{ab}.
\end{equation}
Because we generated the $S$-matrix from the uniform
distribution, such comparisons tend to produce extreme cases
where the coupling of direct channels is too strong. Nevertheless a
general tendency can be clearly seen; when the generalized
transmission coefficient is used, the elastic channel is overestimated
and the inelastic channel is underestimated. The impact of EW
transformation is large, when there are a few channels open
(e.g. Fig.~\ref{fig:modtrans} (a)), and the direct cross sections are
large. Under such circumstances the approximated method to calculate
the cross section by employing the generalized transmission
coefficients leads to incorrect answers.

\begin{figure}
  \begin{center}
    \resizebox{0.9\columnwidth}{!}{\includegraphics{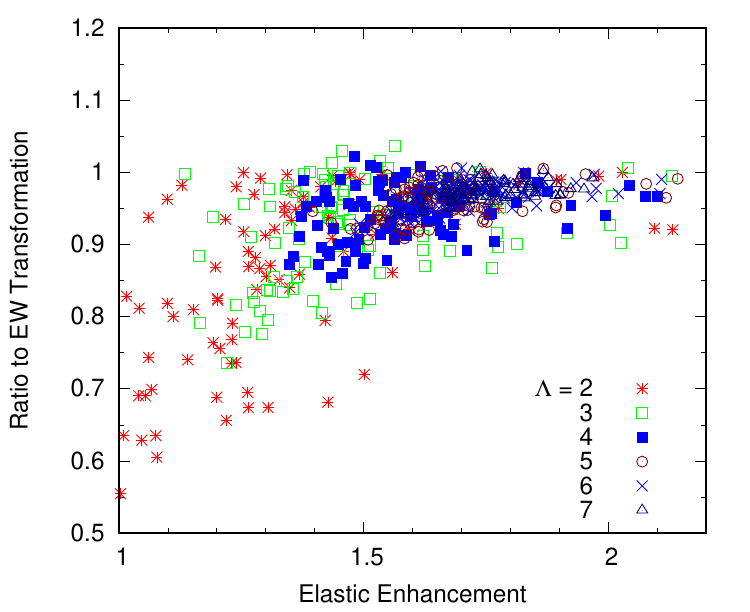}}
  \end{center}
  \caption{Ratio of calculated inelastic scattering cross section with the 
    generalized transmission coefficient calculations to the EW transformation case,
    as a function of the elastic enhancement factor $W_a$.}
  \label{fig:modtranslimit}
\end{figure}

The underestimation in the inelastic channels decreases as the number 
of channels $\Lambda$ increases. That said, we expect that the
approximation with the generalized transmission coefficients works
well at the strong absorption limit, where the elastic enhancement
factor $W_a$ is 2 \cite{Kawano15b}. In our Monte Carlo technique, 
$W_a$ is approximately given by
\begin{equation}
W_a \simeq \sigma_{aa} / \frac{T_a'}{\sum_c T_c'} ,
\end{equation}
where $\sigma_{aa}$ is the compound elastic scattering cross
section. Figure~\ref{fig:modtranslimit} shows the inelastic channel
underestimation as a function of the elastic enhancement. The
underestimation will be very small at the strong absorption limit
($W_a=2$), where the width fluctuation correction to the inelastic
channels fades out due to a large number of open channels. In other
words, the EW transformation is essential when the elastic enhancement
largely changes the inelastic channels.

\subsection{Uncoupled states}

\begin{figure}
  \begin{center}
    \resizebox{0.9\columnwidth}{!}{\includegraphics{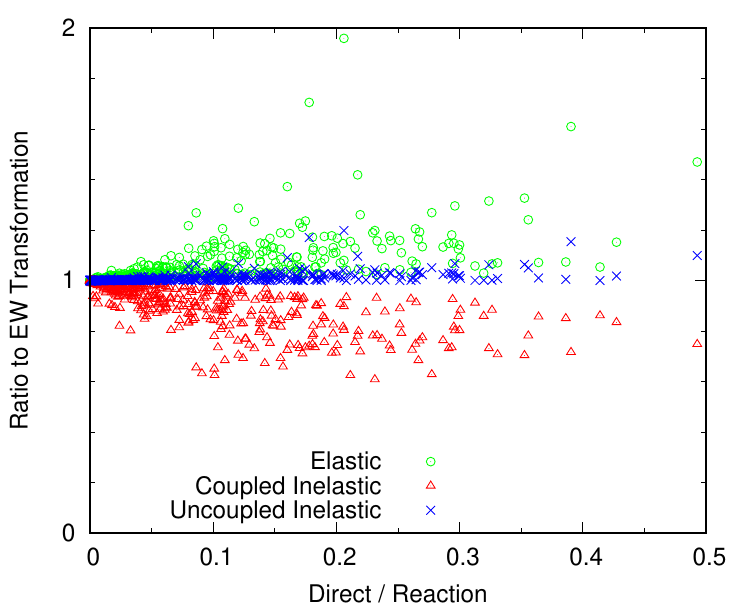}}
  \end{center}
  \caption{Ratio of the cross sections calculated with the generalized transmission
    coefficient calculations to the cross sections calculated with EW transformation
    case, for $\Lambda=3$ and the third channel is uncoupled.}
  \label{fig:modtrans_uncoupled}
\end{figure}

To investigate the uncoupled channel in the EW transformation,
we construct $S$ with $\Lambda=3$ as in
\begin{equation}
  S = \left(
  \begin{array}{ccc}
    S_{aa} & S_{ba} &       \\
    S_{ab} & S_{bb} &       \\
          &        & S_{cc} \\
  \end{array} \right),
\end{equation}
where the channel $c$ is uncoupled to the channels $a$ and $b$.  The
calculated cross sections with the generalized transmission
coefficients are shown by the ratio to the EW transformation in
Fig.~\ref{fig:modtrans_uncoupled}. As opposed to the coupled inelastic
scattering channel, the cross section to the uncoupled channel
increases very slightly, but is almost not influenced by the channel
coupling.  This suggests, in the case of neutron-induced reactions on
deformed nuclei, that the inelastic scattering cross sections will be
enhanced mainly at the expense of the elastic channel, while the
neutron capture and fission cross sections will practically not change.

\subsection{Simulation for Moldauer's estimation}

\begin{figure}
  \begin{center}
    \resizebox{0.9\columnwidth}{!}{\includegraphics{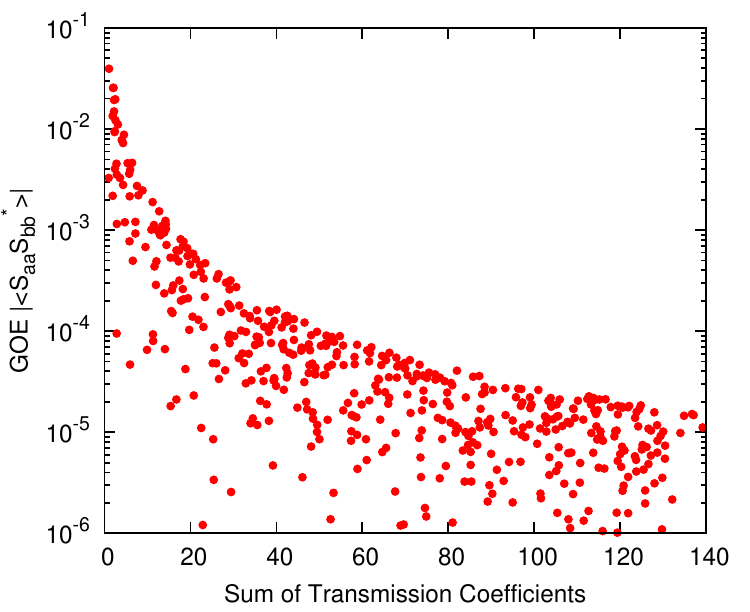}}
  \end{center}
  \caption{Calculated $|\overline{S_{aa} S_{bb}^*}|$ with the GOE triple-integral
    formula for randomly generated $S$-matrix and number of channels.
    The results are shown as a function of $\sum_c T_c$.}
  \label{fig:SabRandom}
\end{figure}

Because the term of $\overline{\tilde{S}_{\alpha\alpha}
  \tilde{S}_{\beta\beta}^*}$ in Eq.~(\ref{eq:backtrans2}) is a
quantity in the diagonalized channel space, we can evaluate this with
the GOE triple-integral of Eq.~(\ref{eq:GOE3int}) whenever $\ave{S}$
is diagonal. We replace $\tilde{S}_{\alpha\alpha}$ by $\ave{S_{aa}}$,
and apply the Monte Carlo technique to calculate $\overline{S_{aa}
  S_{bb}^*}$ by sampling the diagonal $S$-matrix, as well as the
number of channels $\Lambda$ that is randomly varied from 2 to 200. We
generated 500 such random $S$-matrices, and the calculated
$|\overline{S_{aa} S_{bb}^*}|$ is shown by the symbols in
Fig.~\ref{fig:SabRandom}. When there are many open channels, $\sum_c
T_c\gg 1$, this term will be negligible.

\begin{figure}
  \begin{center}
    \resizebox{0.9\columnwidth}{!}{\includegraphics{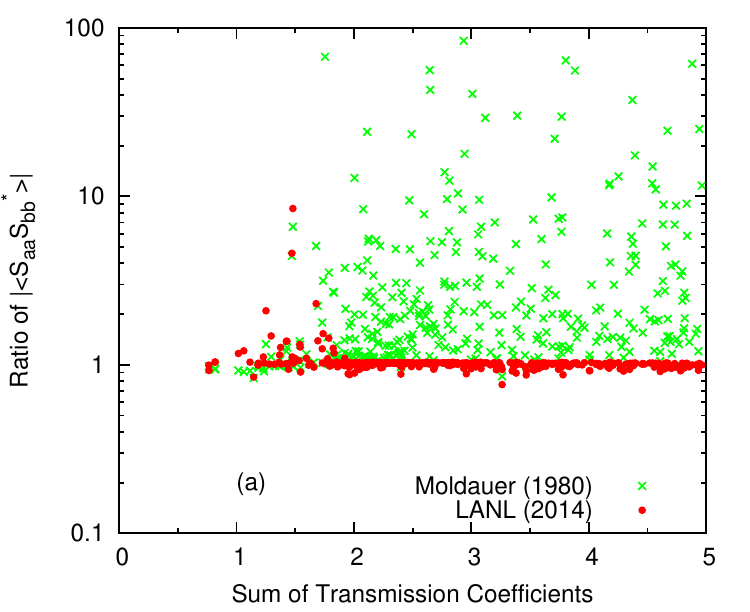}}\\
    \resizebox{0.9\columnwidth}{!}{\includegraphics{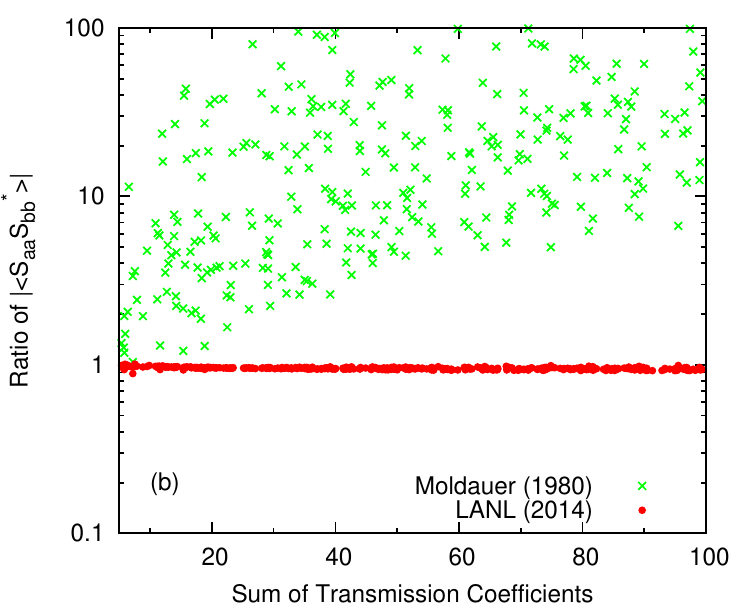}}\\
  \end{center}
  \caption{Comparison of Moldauer's estimate for $|\overline{S_{aa} S_{bb}^*|}$
    given by Eq.~(\ref{eq:Sab}) for various $T_a$ values and channels,
    shown by the ratios to the GOE calculation.
    Two different estimates for the channel degree-of-freedom $\nu$,
    Refs.~\cite{Moldauer80} and \cite{Kawano14}, are used;
    the top panel (a) is for smaller $\sum_c T_c$ case,
    and the bottom panel (b) is for larger $\sum_c T_c$ case.}
   \label{fig:SabRatio}
\end{figure}

Applying two different estimates for $\nu_a$ obtained by Moldauer
\cite{Moldauer80} and at LANL \cite{Kawano14}, Eq.~(\ref{eq:Sab2}) can
be evaluated very easily. Figure~\ref{fig:SabRatio} shows the ratio of
Eq.~(\ref{eq:Sab2}) to the GOE results, using two functional forms for
$\nu_a$.  Since $\overline{S_{aa} S_{bb}^*}$ is complex due to the
factor of $\overline{S_{aa}} \overline{S_{bb}^*}$ in Eq.~(\ref{eq:GOE3int}), the
ratio is taken for the absolute value (the module). It can be seen
clearly that the updated systematics of $\nu_a$ at LANL produces an
excellent agreement with GOE, except for in the very small $\sum_c
T_c$ region, where all statistical models tend to fail
\cite{Kawano14}.

\subsection{Simulation for cross section}

Our next step is to confirm whether Eq.~(\ref{eq:backtrans2}) with the
estimation for $\overline{\tilde{S}_{\alpha\alpha}
  \tilde{S}_{\beta\beta}^*}$ in Eq.~(\ref{eq:Sab2}) is a good
approximation for the actual cross section calculations. To this end,
we calculate the cross sections using the randomly generated
non-diagonal $S$-matrix again, and compare with the reference GOE
cross sections.

\begin{figure}
  \begin{center}
    \resizebox{\columnwidth}{!}{\includegraphics{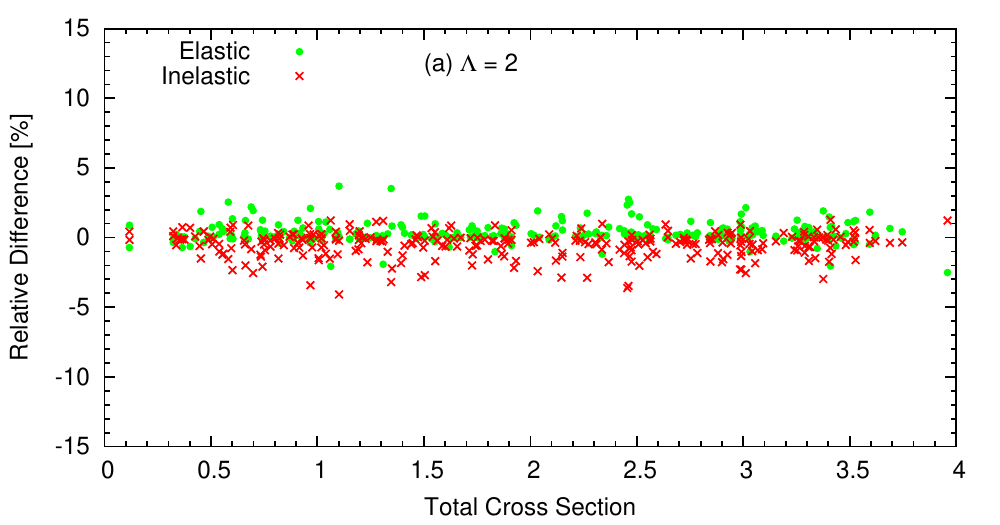}} \\
    \resizebox{\columnwidth}{!}{\includegraphics{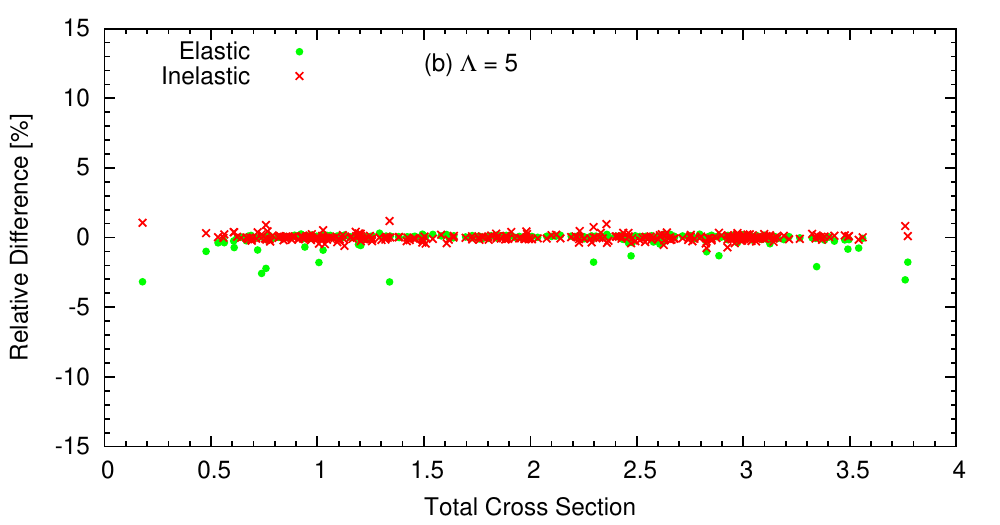}}
  \end{center}
  \caption{Compound elastic and inelastic cross sections calculated
    with randomly sampled $S$-matrix as well as using Moldauer's
    estimate for $|\overline{\tilde{S}_{\alpha\alpha}
      \tilde{S}_{\beta\beta}^*|}$, as a function of the dimensionless
    total cross section.  The results are shown by the deviation from
    the GOE results.  The top panels are for the two channels case,
    and the bottom panels are for the five channels.}
  \label{fig:ewtsim}
\end{figure}

The calculated cross sections for the compound elastic and inelastic
channels are shown by the deviation from GOE in Fig.~\ref{fig:ewtsim},
as a function of total cross section $\sigma^{\rm T}$.  The standard
deviation is 0.83\% for the $\Lambda=2$ case, and 0.29\% for
$\Lambda=5$. From this comparison, we conclude that Moldauer's model
of Eq.~(\ref{eq:Sab}) with the additional phase factor provides a very
good approximation to the GOE triple-integral formula when the
off-diagonal elements in the $S$-matrix exist.  In reality, because
the actual direct channel coupling is much weaker than our randomly
generated $S$-matrix, and the number of channels tends to be larger,
Eqs~(\ref{eq:backtrans2}) and (\ref{eq:Sab2}) should provide an
excellent alternative procedure to calculate compound reaction cross
sections, leading to almost identical cross sections as the
rigorous GOE formula \cite{Nishioka89}.

\section{Coupled-channels and Hauser-Feshbach model in a realistic case}
\label{sec:result2}

We now calculated compound cross sections for neutron induced reactions on
$^{238}$U in the fast energy range with the coupled-channels
Hauser-Feshbach code CoH$_3$, and implement the EW transformation as
well as all the necessary formulae given previously. Note that the
intention here is not to provide the best evaluated cross
section, but to study how large the impact of the EW
transformation on actual cross section calculations will be.  Albeit
it is redundant, we summarize here the procedure of cross section
calculation including the EW transformation as a practical recipe for
applications.

\begin{itemize}
\item For a given total spin and parity $J\Pi$, solve the coupled-channels
      equation. The coupled-channels $S$-matrix is converted into
      $P$-matrix by Eq.~(\ref{eq:Pmatrix}), then diagonalized by
      $UPU^\dag$ to obtain the eigenvalues $p_\alpha$ and the
      eigenvector $U$. We also need the diagonalized $S$-matrix,
      $\tilde{S} = USU^T$.

\item Calculate the transmission sum for all open channels as
\begin{equation}
  T = \sum_\alpha p_\alpha + \sum_k T_k(\mbox{uncoupled}) .
\end{equation}

\item Calculate the channel cross section matrix in the transformed space
\begin{equation}
  \sigma_{\alpha\beta} = \frac{p_\alpha p_\beta}{T} W_{\alpha\beta} ,
\end{equation}
      where the width fluctuation factor $W_{\alpha\beta}$ is given by
      Eq.~(\ref{eq:MoldauerW}).

\item For a set of coupled levels, given a fixed set of
      incoming ($a$) and outgoing ($b$) channels, sum over $a$ and $b$
      when $a \in$ (ground state), and $b \in$ (ground or excited
      state). Summation $\alpha$ and $\beta$ runs over all the
      diagonal space, and calculate the cross section as in
      Eq~.(\ref{eq:backtrans2}) with Eqs.~(\ref{eq:channelsig}) and
      (\ref{eq:Sab2}).

\item For uncoupled levels, run $a$ over the channels that belong to
      the ground state. The cross section is given by Eq.~(\ref{eq:uncoupled}).
\end{itemize}

We employed the dispersive coupled-channels optical potential by
Soukhovitskii et al. \cite{Soukhovitskii05}, with the deformation
parameters of $\beta_2= 0.214$, $\beta_4 = 0.00931$, and $\beta_6 =
-0.0148$ taken from the Finite Range Droplet Model \cite{Moller95}.
We coupled five levels in the ground state rotational band, $0^+$,
$2^+$, $4^+$, $6^+$, and $8^+$. Although direct inelastic scattering
to the vibrational bands can be observed, we consider them as
uncoupled levels to simplify the calculations, otherwise a different
optical model would be needed.

The photon strength function is calculated with the Giant Dipole
Resonance (GDR) model with the parameters of Ullmann et
al. \cite{Ullmann14}. The level density of $^{239}$U is calculated
with Gilbert and Cameron's composite formula \cite{Gilbert65,
  Kawano06}, and the level density parameter is slightly adjusted to
reproduce the average resonance spacing of $D_0=20.26\pm 0.72$~eV
\cite{Mughabghab06}. The fission barrier parameters are taken from
Iwamoto's study \cite{Iwamoto07}, and adjusted to roughly
reproduce the evaluated fission cross section at 1~MeV in ENDF/B-VII
\cite{ENDF7}.  Note that the fission channel is not important, since
we are mainly interested in the cross sections in the sub-threshold
fission region.

\begin{figure*}
  \begin{center}
  \begin{tabular}{cc}
    \resizebox{0.4\textwidth}{!}{\includegraphics{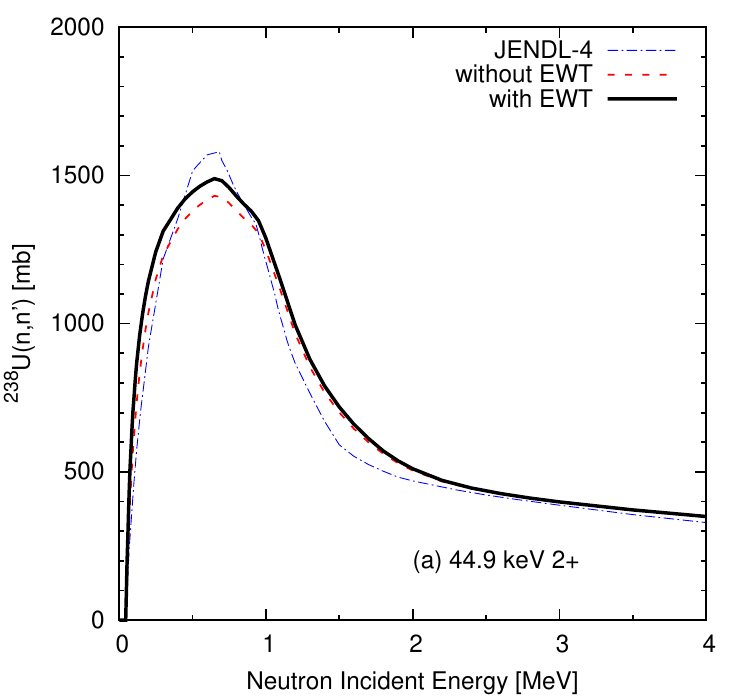}}&
    \resizebox{0.4\textwidth}{!}{\includegraphics{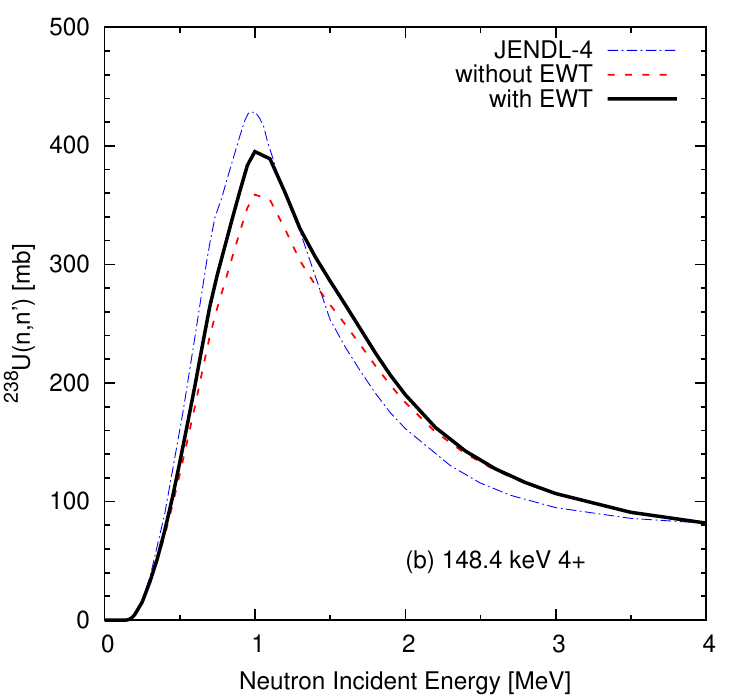}}\\
    \resizebox{0.4\textwidth}{!}{\includegraphics{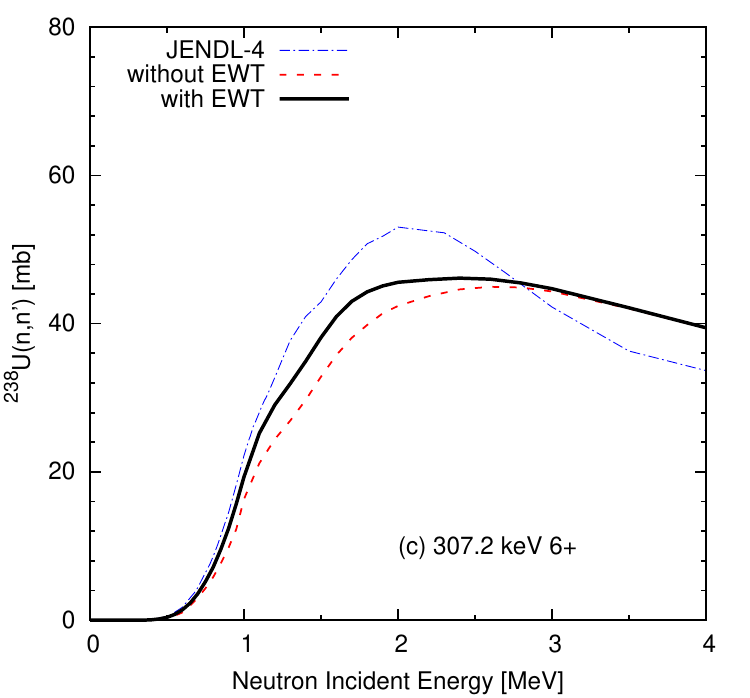}}&
    \resizebox{0.4\textwidth}{!}{\includegraphics{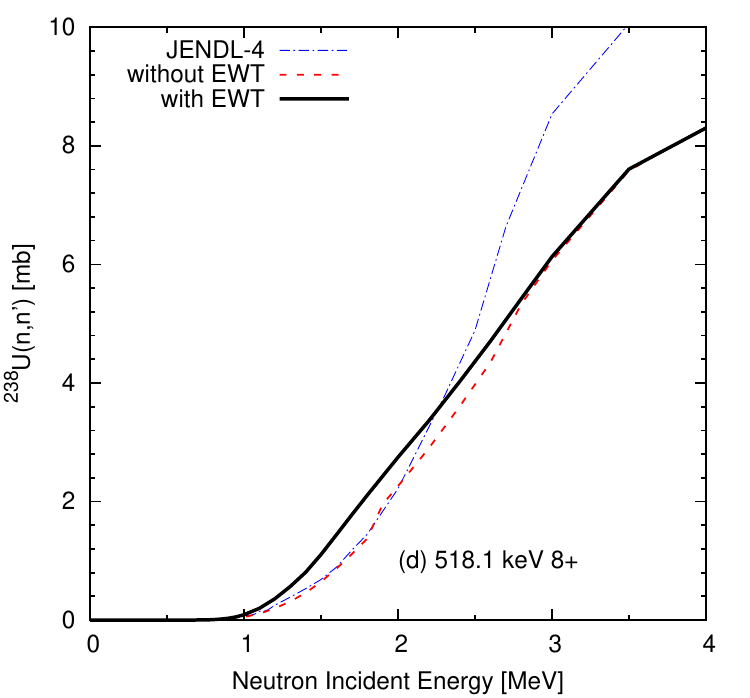}}\\
  \end{tabular}
  \end{center}
  \caption{Calculated $^{238}$U(n,n') reaction cross sections with the
    EW transformation (solid curves) compared with the modified transmission
    calculation (dashed curves), as well as with the evaluated cross sections
    in JENDL-4 (dot-dashed curves).}
  \label{fig:u238inel}
\end{figure*}

Figure~\ref{fig:u238inel} shows the comparison of calculated inelastic
scattering cross sections for the $2^+$, $4^+$, $6^+$, and $8^+$
states.  The dashed curves are calculated with the generalized
transmission coefficients as in Eq.~(\ref{eq:HFmodT}). We also depict
the evaluated cross sections in JENDL-4 \cite{Iwamoto07, JENDL4} for
comparison, since these cross sections were calculated with a similar
optical model with the coupled-channels Hauser-Feshbach code, CCONE
\cite{Iwamoto07}, in which the generalized transmission coefficients are
adopted. The solid curves are the result of EW transformation. The
transformation always increases the inelastic scattering cross section
to the level that has the direct component, which we already observed
in Fig.~\ref{fig:modtrans} in the randomly generated $S$-matrix
model. Because the compound formation cross section $\sigma^{\rm CN}$
remains the same, the increase in the inelastic channels
reduces the enhancement in the compound elastic channel.
However, the reduction in the elastic scattering cross section is not
so visible, since the shape elastic scattering $\sigma_{\rm SE}$
dominates the elastic channel in this energy range.

\begin{figure}
  \begin{center}
    \resizebox{0.9\columnwidth}{!}{\includegraphics{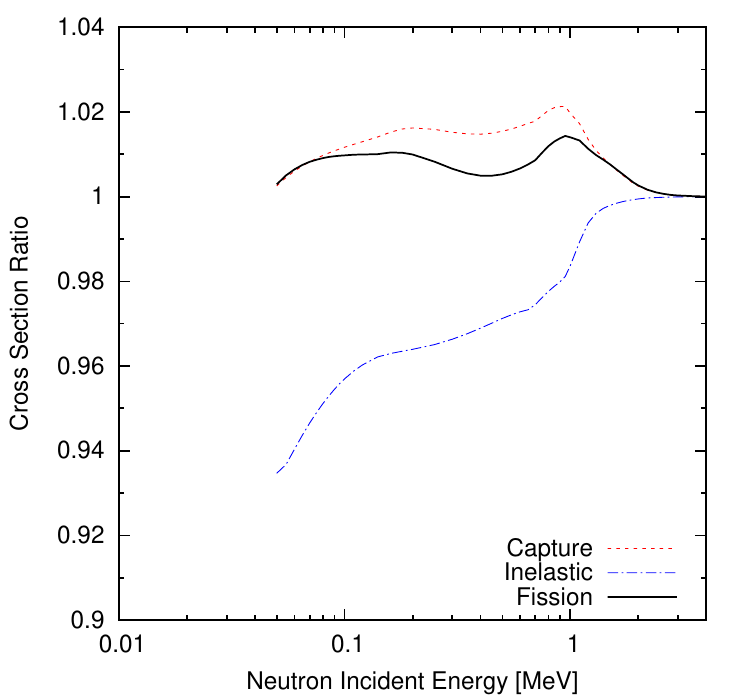}}
  \end{center}
  \caption{Ratios of calculated capture, total inelastic and fission
    cross sections without EW transformation to the EW cases.}
  \label{fig:u238crx}
\end{figure}

The calculated capture, total inelastic, and fission cross sections
are shown in Fig.~\ref{fig:u238crx}, as a ratio of the EW
transformation case to the generalized transmission case. The total
inelastic scattering includes both the coupled and uncoupled
levels. As we already saw in Fig.~\ref{fig:modtrans_uncoupled}, the
generalized transmission calculation gives slightly larger cross
sections for the uncoupled capture and fission channels. However, the
change in these cross sections are less than 2\%, while uncertainties
in the calculated capture and fission cross sections are much larger
in general.

The ratios approach to unity as the neutron incident energy increases,
and the impact of the EW transformation disappears above a few
MeV. Above that energy, the compound elastic scattering cross section
can be basically ignored, because there are many open channels.
Under such circumstances the Hauser-Feshbach theory is justified, and
the cross sections can be calculated without the EW transformation.

\section{Conclusion}

An exact formula for the width fluctuation corrected Hauser-Feshbach
cross section, in which directly coupled channels are involved, is
used to perform the statistical model calculation based on Gaussian
Orthogonal Ensemble (GOE) in the diagonalized space --- the so-called
Engelbrecht-Weidenm\"{u}ller (EW) transformation.  Nishioka,
Weidenm\"{u}ller, and Yoshida \cite{Nishioka89} obtained an equivalent
expression of the fluctuation cross section without the
diagonalization procedure.  Nevertheless, the latter has not been
employed in practical cross section calculations, due to the
complexity both in the formula itself and technical difficulties in
applying actual cases. To overcome this problem, we have developed an
approximated method, which produces almost identical cross sections as
the theory of Nishioka et al., and is feasible to compute cross
sections in realistic cases without any of the difficulties the GOE
inherently possesses. The method combines Moldauer's approximation
\cite{Moldauer75b} with a simple relation between the channel
degree-of-freedom and the optical model transmission coefficient,
recently obtained by a GOE numerical study at LANL \cite{Kawano14}.

We have confirmed the Moldauer's approximation for the first time by
our Monte Carlo approach, and found that an extra phase factor should
be included when ${\rm Im}(\overline{S}_{\alpha\alpha}\overline{S}_{\beta\beta})\ne 0$.  The
method was applied to the description of neutron induced reactions on
$^{238}$U target in the fast energy range, where the elastic and
inelastic scattering, the radiative neutron capture and the fission
channels are relevant. We demonstrated that the EW transformation
indeed increases the calculated inelastic scattering cross sections,
while modest changes were seen in the uncoupled channels, including
the fission and capture cross sections.  We concluded that
conventional methods calculating the Hauser-Feshbach theory by
adopting the generalized (direct cross section eliminated)
transmission coefficients lead to underestimation of the inelastic
scattering cross sections, when the direct channels are strongly
coupled. This underestimation decreases as the number of open channels
increases. We believe this technique should be adopted by existing
Hauser-Feshbach codes, leading to more accurate predictions of the
scattering cross sections on collective nuclei.

\section*{Acknowledgment}

One of the authors (TK) carried out this work under the auspices of
the National Nuclear Security Administration of the U.S. Department of
Energy at Los Alamos National Laboratory under Contract
No. DE-AC52-06NA25396.

\end{document}